\documentclass[12pt]{article}
\usepackage{amssymb}
\usepackage{graphicx}

\begin{document}

\begin{titlepage}
\begin{center}

\vspace*{10mm}

{\LARGE\bf
Tadpole Resummations in String Theory}

\vspace*{20mm}

{\large
Noriaki Kitazawa
}
\vspace{6mm}

{\it
Department of Physics, Tokyo Metropolitan University,\\
Hachioji, Tokyo 192-0397, Japan\\
e-mail: kitazawa@phys.metro-u.ac.jp
}

\vspace*{15mm}

\begin{abstract}
While R-R tadpoles should be canceled for consistency, string models
with broken supersymmetry generally have uncanceled NS-NS tadpoles.
Their presence signals that the background does not solve the field
equations, so that these models are in ``wrong'' vacua. In this
letter we investigate, with reference to some prototype examples,
whether the true values of physical quantities can be recovered
resumming the NS-NS tadpoles, hence by an approach that is related
to the analysis based on String Field Theory by open-closed duality.
We show that, indeed, the positive classical vacuum energy of a
D$p$-brane of the bosonic string is \emph{exactly} canceled by the
negative contribution arising from tree-level tadpole resummation,
in complete agreement with Sen's conjecture on open-string tachyon
condensation and with the consequent analysis based on String Field
Theory. We also show that the vanishing classical vacuum energy of
the SO$(8192)$ unoriented bosonic open-string theory does not
receive any tree-level corrections from the tadpole resummation.
This result is consistent with the fact that this (unstable)
configuration is free from tadpoles of massless closed-string modes,
although there is a tadpole of the closed string tachyon. The
application of this method to superstring models with broken
supersymmetry is also discussed.
\end{abstract}

\end{center}
\end{titlepage}

\section{Introduction}
\label{introduction}

String models with tension in the TeV region \cite{Antoniadis:1990ew}
are an exciting possibility for physics beyond the Standard Model
 (for a review, see refs.~\cite{Antoniadis:2004wm,Antoniadis:2007uz}).
This scenario was made concrete by the modern understanding of open strings
and orientifold compactifications \cite{orient}, and by many important
subsequent developments (for a review, see ref.~\cite{Blumenhagen:2006ci}).
In this framework,
the Higgs doublet responsible for the electroweak symmetry breaking
is typically identified with some massless open-string mode on
D$3$-branes, that can acquire a negative mass squared via radiative
corrections in models without
supersymmetry \cite{Antoniadis:2000tq,Kitazawa:2006if}. The scale of
electroweak symmetry breaking is essentially determined by the
string tension (and/or by the radii of the compact space), and this
scenario can be a natural dynamical setting where string effects
play a direct role for Particle Physics.

In string models with broken supersymmetry, however, there is in
general a vexing difficulty, the so called NS-NS tadpole problem.
The existence of tadpoles in the NS-NS closed-string sector signals
that the assumed background metric and field configuration
(typically flat spacetime with vanishing background fields)
\emph{is not} a solution of String Theory. On the other hand, it is
quite difficult to construct string models with more general
background metrics or non-trivial fields, so that a prescription to
cure this problem proposed in
refs.~\cite{Fischler:1986ci,Fischler:1986tb,Das:1986dy} is also not
easy to implement. The actual difficulty manifests itself via the
emergence of infrared divergences in loop calculations of open
string amplitudes.

Tadpole resummation as a possible way to overcome this problem was
proposed in ref.~\cite{Dudas:2004nd}, where several examples of field
theories defined in ``wrong" vacua were discussed. Indeed, barring
convergence issues and other subtleties, the correct value of the
vacuum energy can generically be recovered by the procedure of
tadpole resummation. This procedure is nonetheless quite
complicated, since it requires that one add up all tree-level
diagrams involving tadpole contributions.

In this letter we propose a concrete method to implement this
procedure in String Theory, showing that tadpole resummations
actually lead to the correct answer for open-string tachyon
condensation. To this end, we combine the boundary state formalism
with some information drawn from the low-energy dynamics of branes.
In the next section the boundary state formalism for D-branes and
O-planes is thus briefly reviewed. In Section \ref{resum_Dp} we
study the vacuum energies (tensions) of D$p$-branes for the bosonic
string \footnote{These D$p$-branes do not carry RR charges, but are
nonetheless characterized as being locations of the endpoints of
open strings.} in flat $26$-dimensional spacetime as a first simple
example. These D$p$-branes have generally tadpoles for dilaton,
graviton and tachyon modes. The conjecture of open string tachyon
condensation \cite{Sen:1998sm} claims that they should decay to the
vacuum, so that the actual vacuum energy should vanish. This
conjecture has received strong support from String Field Theory
\cite{string-field-thery}, although the resulting mechanism appears
rather complicated and rests crucially on the contributions of open
string massive modes. Here we show that, rather remarkably, the
phenomenon can be understood in somewhat simpler terms: in the dual
closed channel a \emph{negative} contribution originating from
tree-level tadpole resummations \emph{exactly cancels} the positive
classical vacuum energy of the D$p$-brane. Section \ref{resum_D25}
is devoted to a similar analysis of tadpole resummations for the
D$25$-branes of the SO$(8192)$ open bosonic string
\cite{Douglas:1986eu,Marcus:1986cm,Weinberg:1987ie}. The absence of
massless dilaton and graviton tadpoles makes somehow this
D$25$-brane -- O$25$-plane system a solution of String Theory,
albeit an unstable one. We show that, indeed, in this case
tree-level tadpole resummations do not produce any correction to the
D$25$-brane tension, despite the presence of a tadpole for the
tachyon. The last section is devoted to a brief discussion of the
limitations of the method (originating from the actual neglect of
the gravitational back reaction) and of its application to
superstring models with broken supersymmetry.

\section{Boundary states in the bosonic string}
\label{boundary_states}

In superstrings, D-branes can be conveniently described by boundary
states for the closed string in the world-sheet theory (for a
review, see ref.~\cite{DiVecchia:1999rh}). The technique also applies
for the D-branes of the bosonic string, that despite the lack of a
RR charge, bear strong similarities to their supersymmetric
counterparts. For the $26$-dimensional bosonic string in flat
spacetime, a D$p$-brane boundary state at the origin, $\vert B_p
\rangle$, satisfies the conditions
\begin{eqnarray}
 \partial_\tau X^\alpha(\sigma,\tau=0) \vert B_p \rangle &=& 0\, ,
   \qquad \alpha = 0,1,\cdots,p\, ,
\\
 X^i(\sigma,\tau=0) \vert B_p \rangle &=& 0\, ,
   \qquad i = p+1,p+2,\cdots,25\, ,
\end{eqnarray}
where $X^\mu(\sigma,\tau)$ is a closed-string coordinate. The
boundary state can be explicitly expressed as a coherent state built
from the string oscillators,
\begin{equation}
 \vert B_p \rangle = \vert B_p^X \rangle \vert B^{\rm gh} \rangle,
\end{equation}
\begin{eqnarray}
 \vert B_p^X \rangle &=&
  N_p
  \delta^{d_\perp}(\hat{x})
   \exp\left(
        - \sum_{n=1}^\infty {1 \over n}
          \alpha_{-n}^\mu S_{\mu\nu} {\tilde \alpha}_{-n}^\nu
        \right) \vert 0 \rangle,
\\
 \vert B^{\rm gh} \rangle &=&
   \exp\left(
        \sum_{n=1}^\infty
         \left(
          c_{-n} {\tilde b}_{-n} - b_{-n} {\tilde c}_{-n}
         \right)
        \right)
        \vert 0 \rangle_{\rm gh},
\end{eqnarray}
where $d_\perp \equiv d-(p+1)$, $S_{\mu\nu} \equiv
(\eta_{\alpha\beta}, -\delta_{ij})$,
 the spacetime signature is ``mostly plus'', and
\begin{equation}
 X^\mu =
  \hat{x}^\mu + \alpha' p^\mu \tau
  + i \sqrt{{{\alpha'} \over 2}}
    \sum_{n \ne 0} {1 \over n}
     \left(
      \alpha_n^\mu \ e^{-i(\tau-\sigma)n}
      + {\tilde \alpha}_n^\mu \ e^{-i(\tau+\sigma)n}
     \right),
\end{equation}
\begin{eqnarray}
 b_- = \sum_{n=-\infty}^\infty b_n e^{-i(\tau-\sigma)n},
 &&
 b_+ = \sum_{n=-\infty}^\infty {\tilde b}_n e^{-i(\tau+\sigma)n},
\\
 c_- = \sum_{n=-\infty}^\infty c_n e^{-i(\tau-\sigma)n},
 &&
 c_+ = \sum_{n=-\infty}^\infty {\tilde c}_n e^{-i(\tau+\sigma)n}.
\end{eqnarray}
The $bc$-ghost contribution is determined so that the full boundary
state is BRST invariant, while the normalization constant $N_p$ for
one D$p$-brane is
\begin{equation}
 N_p \equiv {{T_p} \over 2},
 \qquad
 T_p \equiv {\sqrt{\pi} \over {2^{(d-10)/4}}}
            ( 4 \pi^2 \alpha' )^{{d-2p-4} \over 4}
\end{equation}
with $d=26$ and the $26$-dimensional Planck constant $\kappa=1$, so
that the single dilaton and graviton tadpole couplings in the
Einstein frame are reproduced \cite{DiVecchia:1997pr}. For a
collection of $n$ coincident D$p$-branes the normalization factor
should be multiplied by $n$. The amplitudes for single dilaton or
graviton emission are
\begin{eqnarray}
 A_{\rm dilaton} &=& A^{\mu\nu} \epsilon^{(\phi)}_{\mu\nu}
  = T_p V_{p+1} {{d-2p-4} \over {2\sqrt{d-2}}},
\\
 A_{\rm graviton} &=& A^{\mu\nu} \epsilon^{(h)}_{\mu\nu}
  = - T_p V_{p+1} \eta^{\alpha\beta} \epsilon^{(h)}_{\alpha\beta},
\end{eqnarray}
\begin{equation}
 A^{\mu\nu} \equiv
  \langle 0;k \vert \alpha_1^\mu \tilde{\alpha}_1^\nu \vert B_p^X \rangle
  = - {{T_p} \over 2} V_{p+1} S^{\mu\nu},
\end{equation}
where $V_{p+1}$ is the D$p$-brane world volume and
$\epsilon^{(\phi)}_{\mu\nu}$ and $\epsilon^{(h)}_{\mu\nu}$ are
projection and polarization tensors for the dilaton and the
graviton, respectively. These amplitudes are also determined by the
effective action for a D$p$-brane in the Einstein frame,
\begin{equation}
\label{DBI}
 S_{{\rm D}p} = - T_p \int d^{p+1} \xi \
  e^{- {{d-2p-4} \over {2\sqrt{d-2}}} \phi}
  \sqrt{- \det g_{\alpha\beta}},
\end{equation}
where we are ignoring both the $B$-field and the gauge field on the
D$p$-brane, for simplicity. With this normalization factor,
the open string one-loop vacuum amplitudes on D$p$-branes are
simply obtained as
\begin{equation}
\label{cylinder}
 {\cal A}_p =
 {1 \over {2!}} \langle B_p \vert D \vert B_p \rangle
  = {1 \over {2!}} V_{p+1} N_p^2 \Delta_p,
\end{equation}
\begin{eqnarray}
 \Delta_p &\equiv& {{\pi \alpha'} \over 2} \int_0^\infty ds
                   \int {{d^d_\perp p} \over {(2\pi)^{d_\perp}}}
                   e^{-{{\pi \alpha'} \over 2} p_\perp^2 s}
                   {1 \over {\eta(is)^{24}}}
\\
               &=& {{\pi \alpha'} \over 2} \int_0^\infty ds
    {1 \over {(2 \pi^2 \alpha' s)^{d_\perp/2}}}
                   {1 \over {\eta(is)^{24}}},
\end{eqnarray}
where
\begin{equation}
 D \equiv {{\alpha'} \over {4\pi}}
          \int_0^\infty dt \int_0^{2\pi} d\varphi \
           z^{L_0} {\bar z}^{{\tilde L}_0}
\end{equation}
is the closed string propagator operator, with
$z=e^{-t}e^{i\varphi}$. The factor $1/2!$ in eq.~(\ref{cylinder})
reflects the fact that the amplitude is of second order in
the tadpole insertion.

The boundary state of the O-plane, the crosscap state, is very
similar. For the O$25$-plane, the crosscap state should satisfy the condition
\begin{equation}
 X^\mu(\sigma,\tau) \vert C_{25} \rangle
  = X^\mu(\sigma,\pi-\tau) \vert C_{25} \rangle,
  \qquad \mu = 0,1,\cdots,25.
\end{equation}
The explicit form of the crosscap state is
\begin{equation}
 \vert C_{25} \rangle = \vert C_{25}^X \rangle \vert C^{\rm gh} \rangle,
\end{equation}
\begin{eqnarray}
 \vert C_{25}^X \rangle &=&
  {\tilde N}_{25}
   \exp\left(
        - \sum_{n=1}^\infty (-1)^n {1 \over n}
          \alpha_{-n}^\mu \eta_{\mu\nu} {\tilde \alpha}_{-n}^\nu
        \right) \vert 0 \rangle,
\\
 \vert C^{\rm gh} \rangle &=&
   \exp\left(
        \sum_{n=1}^\infty (-1)^n
         \left(
          c_{-n} {\tilde b}_{-n} - b_{-n} {\tilde c}_{-n}
         \right)
        \right)
        \vert 0 \rangle_{\rm gh}.
\end{eqnarray}
The normalization constant ${\tilde N}_{25}$ is determined as
${\tilde N}_{25} = 2^{13} N_{25}$ in the same way as $N_p$. The
Klein bottle and M\"obius strip amplitudes read
\begin{eqnarray}
\label{klein}
 {\cal K}
 &=& {1 \over {2!}}\langle C_{25} \vert D \vert C_{25} \rangle
 = {1 \over {2!}} V_{26} {\tilde N}_{25}^2 \Delta_{25}\, ,
\\
\label{moebius}
 {\cal M} &=&
   {1 \over {2!}} \left(
                   \langle B_{25} \vert D \vert C_{25} \rangle
                   +  \langle C_{25} \vert D \vert B_{25} \rangle
                  \right)
 = V_{26} N_{25} {\tilde N}_{25} {\tilde \Delta}_{25}\, ,
\end{eqnarray}
where
\begin{equation}
 {\tilde \Delta}_{25}
  \equiv - {{\pi \alpha'} \over 2}
           \int_0^\infty ds {1 \over {\hat{\eta}(is+1/2)^{24}}}\, .
\end{equation}
>From eq.~(\ref{cylinder}) with $p=25$,
and eqs.~(\ref{klein}) and (\ref{moebius}),
one can see that $n=2^{13}=8192$ D$25$-branes,
with the normalization factor $n \times N_{25}$ for D$25$-brane
boundary state, are necessary and sufficient
to cancel tadpoles in the unoriented bosonic closed string theory
\cite{Douglas:1986eu,Marcus:1986cm,Weinberg:1987ie}.

\section{Tadpole resummations on D$p$-branes}
\label{resum_Dp}

At the classical level, the vacuum energy density on a D$p$-brane
coincides with its tension, $\Lambda_p^{\rm cl} = T_p$, and can be
read from eq.~(\ref{DBI}). The open string one-loop correction to
the vacuum energy is given by eq.~(\ref{cylinder}), and contains
divergences due to the tadpoles of dilaton, graviton and tachyon in
$\Delta_p$. These tadpole contributions can be exhibited expanding
the integrand of $\Delta_p$ for large value of $s$, as
\begin{equation}
 \Delta_p \rightarrow
  {{\pi \alpha'} \over 2}
   \int^\infty ds
    {1 \over {(2 \pi^2 \alpha' s)^{d_\perp/2}}}
     \left( e^{2 \pi s} + 24 + {\cal O}(e^{-2 \pi s}) \right).
\end{equation}
The first term within brackets is the contribution of tachyon
tadpoles while the second is the overall contribution from massless
dilaton and graviton tadpoles. These divergences can in principle be
regularized via an ``infrared'' cutoff on $s$ (or ``ultraviolet'',
from the open-channel perspective), and we shall do it implicitly in
what follows.

\begin{figure}[t]
\centering
\includegraphics[width=120mm]{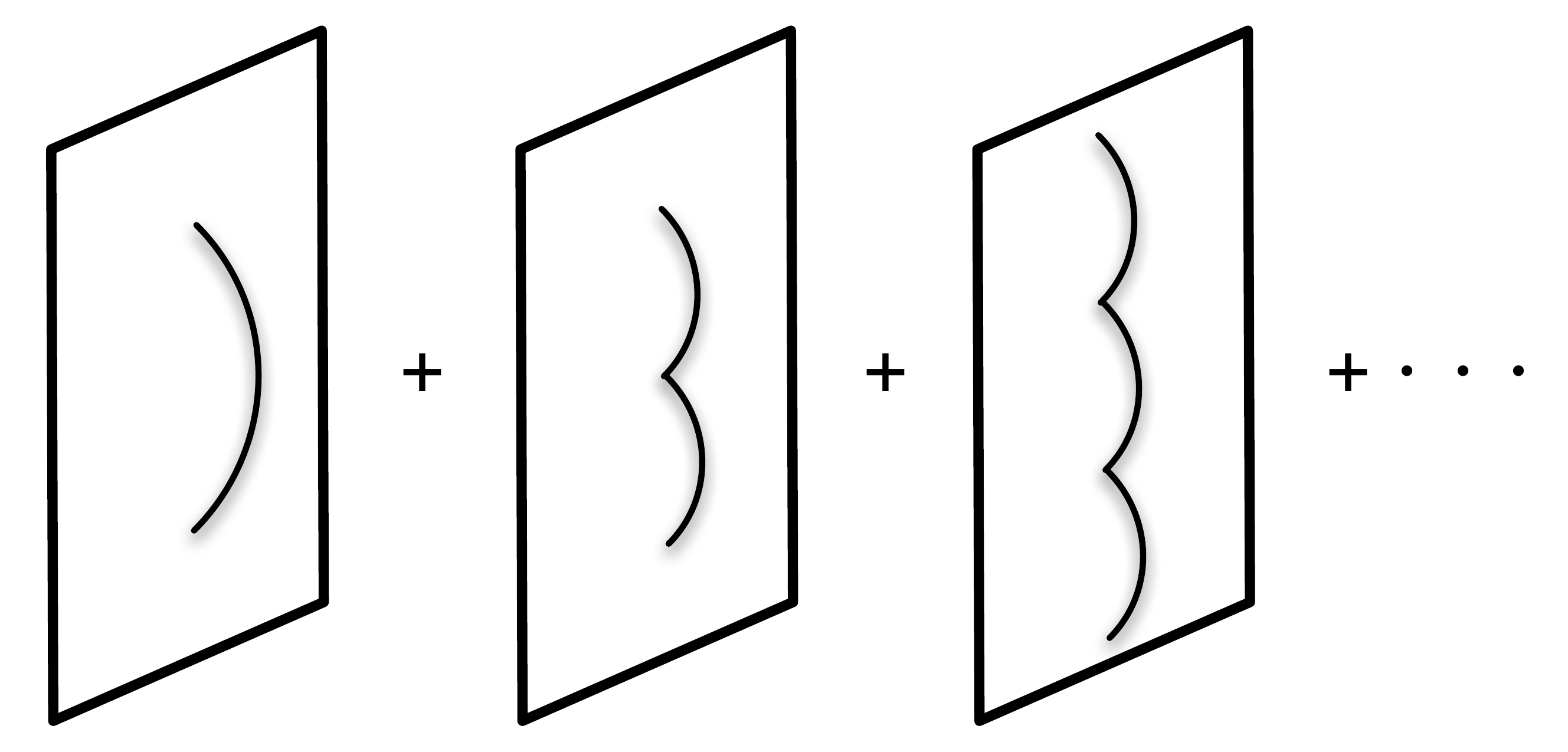}
\caption{Closed string bouncing on a D$p$-brane.}
\label{fig:propagator}
\end{figure}

\begin{figure}[t]
\centering
\includegraphics[width=120mm]{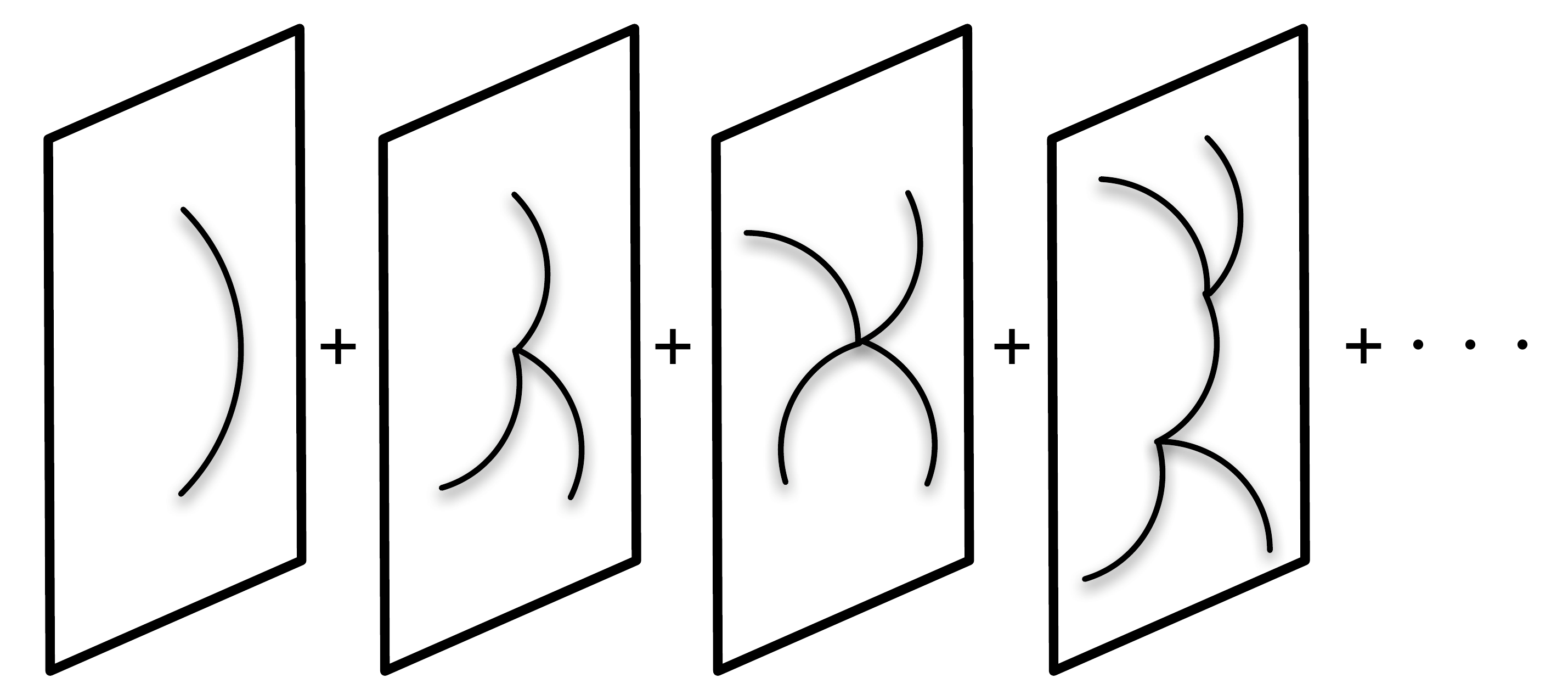}
\caption{Summation of tree-level contributions.}
\label{fig:multipoint}
\end{figure}

It is important to recognize that, in addition to the cylinder amplitude, the simplest contribution in
the closed-string picture, there are many other three-level contributions
with closed string contact interactions,
 as shown in figs.~\ref{fig:propagator} and \ref{fig:multipoint}.
There are also closed-string one-loop and higher-loop contributions, that we neglect in the
present calculations. As we stressed  above, the existence of the contact interactions
between dilaton/graviton and D-branes can be  simply inferred from effective action of eq.~(\ref{DBI}).

The effect of the two point contact interaction, that can be understood as a process in which a
closed string bounces off the D-brane at the origin, can be accounted for inserting the operator
\begin{equation}
 \hat{M}
  \equiv \int d^dx \delta^{d_\perp}(x)
  \vert {\tilde B}_p(x) \rangle
   (-T_p)
  \langle {\tilde B}_p(x) \vert\, ,
\end{equation}
 where
\begin{equation}
 \vert {\tilde B}_p(x) \rangle
  =  \vert {\tilde B}^X_p(x) \rangle \vert B^{\rm gh} \rangle\, ,
\end{equation}
\begin{equation}
 \vert {\tilde B}^X_p(x) \rangle
 \equiv
  {1 \over {T_p}}
  \delta^{p+1}(\hat{x}-x)
  \vert B^X_p \rangle \, .
\end{equation}
The state $\vert {\tilde B}_p(x) \rangle$ is essentially a D$p$-brane
boundary state with a different normalization, with the position of
the closed string on the D$p$-brane fixed at the generic point $x$,
and indeed the integration in the definition of $\hat{M}$ is over
the D$p$-brane world volume. The overall normalization of the operator
$\hat{M}$ is such that the coupling constant of the dilaton two-point
contact interaction coincides with that present in the effective field
theory of eq.~(\ref{DBI}).

%%%%%%%%%%%%%%%%%%%%%%%%%%%%%%%%%%%%%%%%%%%%%%%%%%%%%%%%%%%%%%%%%%%%%%%%%%%%%%
%%%%%%%%%%%%%%%%%%%%%%%%%%%%%%%%%%%%%%%%%%%%%%%%%%%%%%%%%%%%%%%%%%%%%%%%%%%%%%
The ``one-bounce'' contribution to the D$p$-brane vacuum energy of
 fig.~\ref{fig:propagator} is thus
\begin{eqnarray}
 A_1 &=&
  {1 \over {2!}} \langle B_p \vert D \hat{M} D \vert B_p \rangle
\nonumber\\
     &=& {1 \over {2!}} \int d^dx \delta^{d_\perp}(x)
         \langle B_p \vert D \vert {\tilde B}_p(x) \rangle
          (-T_p)
         \langle {\tilde B}_p(x) \vert D \vert B_p \rangle\, .
\end{eqnarray}
The quantity $\langle {\tilde B}_p(x) \vert D \vert B_p \rangle$
can be calculated in the same way as the cylinder amplitude:
\begin{eqnarray}
 \langle {\tilde B}_p(x) \vert D \vert B_p \rangle
 &=& \langle B_p \vert D \vert {\tilde B}_p(x) \rangle
\nonumber\\
 &=& {{N_p^2} \over {T_p}}
     {{\pi \alpha'} \over 2}
     \int_0^\infty ds
      {1 \over {(2 \pi^2 \alpha' s)^{d_\perp/2}}} \
      e^{-{{x^2_\perp} \over {2 \pi \alpha' s}}} \
      {1 \over {\eta(is)^{24}}}.
\end{eqnarray}
Therefore,
\begin{equation}
 A_1 = {1 \over {2!}}
       V_{p+1} N_p^2 \left( {{N_p} \over {T_p}} \right)^2 (-T_p)
       \Delta_p^2.
\end{equation}

The ``two-bounce'', ``three-bounce'' and all the other amplitudes of this type
can then be computed in the same way. The complete ``two-point function''
is then obtained summing all these contributions
as depicted in fig.~\ref{fig:propagator}.
\begin{eqnarray}\label{2ptfinal}
 A^{(2)}_p
  &=& {1 \over {2!}}
      \left\{
       \langle B_p \vert D \vert B_p \rangle
       + \langle B_p \vert D \hat{M} D \vert B_p \rangle
       + \langle B_p \vert D \hat{M} D \hat{M} D \vert B_p \rangle
       + \cdots
      \right\},
\nonumber\\
 &\equiv& {1 \over {2!}} \langle B_p \vert D_M \vert B_p \rangle
\nonumber\\
 &=& {1 \over {2!}} V_{p+1} N_p^2
     {{\Delta_p} \over {1+T_p (N_p/T_p)^2 \Delta_p}}\, .
\end{eqnarray}
The result is a geometric series involving the regularized
$\Delta_p$. Notice that, in the limit that the ``infrared''
regulator for $\Delta_p$ is removed, a simple finite value obtains
for the two-point function:
\begin{equation}
 A^{(2)}_p = {1 \over {2!}} V_{p+1} T_p\, ,
\end{equation}
so that the corresponding contribution to the vacuum energy density
on a D$p$-brane is
\begin{equation}
 \Lambda_p^{(2)}
  = - {{A^{(2)}_p} \over {V_{p+1}}} = - {1 \over {2!}} T_p\, .
\end{equation}
%
%%%%%%%%%%%%%%%%%%%%%%%%%%%%%%%%%%%%%%%%%%%%%%%%%%%%%%%%%%%%%%%%%%%%%%%%%%%%%%
%%%%%%%%%%%%%%%%%%%%%%%%%%%%%%%%%%%%%%%%%%%%%%%%%%%%%%%%%%%%%%%%%%%%%%%%%%%%%%

Let us now turn to the contributions depicted in fig.~\ref{fig:multipoint}.
There is a dilaton three-point contact interaction with the D$p$-brane,
that again can be deduced from eq.~(\ref{DBI}). In the boundary state formalism
this interaction could be represented via an operator acting on three states:
\begin{equation}
 \hat{M}^{(3)} \equiv
 {1 \over {3!}}
 T_p \left( {{N_p} \over {T_p}} \right)^3
   \int d^dx \delta^{d_\perp}(x)
   \vert {\tilde B}_p(x) \rangle
   \vert {\tilde B}_p(x) \rangle
   \vert {\tilde B}_p(x) \rangle.
\end{equation}
The legs of the resulting ``three-point function'' should be \emph{full}
``two-point functions'', namely,
\begin{equation}
 A^{(3)}_p = {1 \over {3!}}
 T_p \int d^dx \delta^{d_\perp}(x)
   \langle B_p \vert D_M \vert {\tilde B}_p(x) \rangle
   \langle B_p \vert D_M \vert {\tilde B}_p(x) \rangle
   \langle B_p \vert D_M \vert {\tilde B}_p(x) \rangle,
\end{equation}
where $1/3!$ is a symmetry factor
\footnote{Its origin is as follows:
the vertex carries a $1/3!$,
the third order of the tadpole insertion brings about one more $1/3!$,
while the number of contractions gives rise to a factor $3!$.}.

A straightforward calculation gives
\begin{equation}
 \langle B_p \vert D_M \vert {\tilde B}_p(x) \rangle
 = {1 \over {1 + T_p (N_p/T_p)^2 \Delta_p}}
   \langle B_p \vert D \vert {\tilde B}_p(x) \rangle\, ,
\end{equation}
and therefore
\begin{equation}
 A^{(3)}_p
  = {1 \over {3!}} T_p V_{p+1}
    \left(
     {1 \over {1 + T_p (N_p/T_p)^2 \Delta_p}}
      \cdot
     {{N_p^2} \over {T_p}} \Delta_p
    \right)^3
  = {1 \over {3!}} V_{p+1} T_p.
\end{equation}
The contribution to the vacuum energy density on a D$p$-brane is thus
\begin{equation}
 \Lambda_p^{(3)}
  = - {{A^{(3)}_p} \over {V_{p+1}}} = - {1 \over {3!}} T_p.
\end{equation}

One can continue this calculation along similar lines for the
``four-point function'', and in fact to all orders. The end result
is very similar to the example of the scalar field with exponential
potential of ref.~\cite{Dudas:2004nd}:
\begin{equation}
 A^{\rm tree}_p \equiv A^{(2)}_p + A^{(3)}_p +A^{(4)}_p + \cdots
 = V_{p+1} T_p \sum_{n=1}^\infty {1 \over {n(n+1)}}
 = V_{p+1} T_p \, ,
\end{equation}
since the series can be easily shown to add up to one. Therefore,
the total contribution to the vacuum energy density on a D$p$-brane
arising from the tree-level tadpole resummation is simply
\begin{equation}
 \Lambda^{\rm tree}_p = - A^{\rm tree}_p/V_{p+1} = -T_p\, ,
\end{equation}
and exactly cancels the classical energy density
$\Lambda^{\rm cl}_p = T_p$.

\section{Tadpole resummations on a D$25$-brane
in the unoriented SO$(8192)$ theory}
\label{resum_D25}

The tree-level tadpole resummation of the previous section is
essentially governed by the fact that the open string one-loop
vacuum amplitude, $\Delta_p$, contains a divergent massless
contribution. It is therefore interesting to verify that the
contribution vanishes in the absence of massless tadpoles, even if a
tachyon tadpole is present. In this section we thus examine the
tadpole resummation in the SO$(8192)$ theory, where massless
tadpoles are canceled as a result of the cooperative action of the
D$25$-branes and an O$25$-plane.

There are three types of open-string one-loop contributions to the
vacuum energy, associated to the cylinder, M\"obius strip and Klein
bottle amplitudes:
\begin{eqnarray}
 {\cal A} &=&
  {1 \over {2!}} \langle B_{25} \vert D \vert B_{25} \rangle
  = {1 \over {2!}} V_{26} N_{25}^2 \Delta_{25}\, ,
\\
 {\cal M} &=&
  {1 \over {2!}} \left(
                  \langle B_{25} \vert D \vert C_{25} \rangle
                  + \langle C_{25} \vert D \vert B_{25} \rangle
                 \right)
  = {1 \over {2!}}
     \cdot 2 \langle C_{25} \vert D \vert B_{25} \rangle
  = V_{26} N_{25} {\tilde N}_{25} {\tilde \Delta}_{25}\, ,
\\
 {\cal K} &=&
  {1 \over {2!}} \langle C_{25} \vert D \vert C_{25} \rangle
  = {1 \over {2!}} V_{26} {\tilde N}_{25}^2 \Delta_{25}\, .
\end{eqnarray}
The divergences due to the tadpoles emerge from
\begin{eqnarray}
 \Delta_{25} &\rightarrow&
  {{\pi \alpha'} \over 2}
   \int^\infty ds
     \left( e^{2 \pi s} + 24 + {\cal O}(e^{-2 \pi s}) \right)\, ,
\\
 \tilde{\Delta}_{25} &\rightarrow&
  {{\pi \alpha'} \over 2}
   \int^\infty ds
     \left( e^{2 \pi s} - 24 + {\cal O}(e^{-2 \pi s}) \right)\, .
\end{eqnarray}
Notice that the tachyon contributions have the same sign, while the
massless dilaton/graviton contributions have opposite signs, in
these two quantities. One can define two divergent quantities
related, respectively, to the tachyon and dilaton/graviton tadpoles,
as
\begin{eqnarray}
 \Delta_T &\equiv& {{\pi \alpha'} \over 2} \int_0^\infty ds e^{2 \pi s}\, ,
\\
 \Delta &\equiv& {{\pi \alpha'} \over 2} \int_0^\infty ds \ 24\, ,
\end{eqnarray}
that can be regulated introducing an ``infrared'' cutoff on $s$. One
can then write
\begin{eqnarray}
 \Delta_{25} &\rightarrow& \Delta_T + \Delta + \mbox{finite},
\\
 \tilde{\Delta}_{25} &\rightarrow& \Delta_T - \Delta + \mbox{finite}.
\end{eqnarray}

In order to obtain the full ``two-point functions'', ``bounce
effects'' on both D$25$-branes and the O$25$-plane should be
included. The ``bounce effect'' on the O$25$-plane can be accounted
for inserting the operator
\begin{equation}
 \hat{M}_O = \int d^{26}x
             \vert \tilde{C}_{25}(x) \rangle
              (+\tilde{T}_{25})
             \langle \tilde{C}_{25}(x) \vert\, ,
\end{equation}
that is obtained in the same way as the operator $\hat{M}$, where
\begin{equation}
 \vert \tilde{C}_{25}(x) \rangle
  \equiv {1 \over {\tilde{T}_{25}}} \delta^{26}(\hat{x}-x)
  \vert C_{25}(x) \rangle
\end{equation}
and $\tilde{T}_{25} \equiv \tilde{N}_{25} / 2$ is the tension of O$25$-plane.

The full ``two-point function'' with both edges on D$25$-branes and
without O$25$-bounces was already given in eq. (\ref{2ptfinal}), and
is
\begin{equation}
 A^{(2)}_{\rm zero-O}
 = {1 \over {2!}} \langle B_{25} \vert D_M \vert B_{25} \rangle
 = {1 \over {2!}} V_{26} N_{25}^2
   {{\Delta_{25}} \over {1+T_{25} (N_{25}/T_{25})^2 \Delta_{25}}},
\end{equation}
where $N_{25}$ and $T_{25}$ include the number of D$25$-branes, $n=2^{13}=8192$.
Therefore, $\tilde{N}_{25}=N_{25}$ and $\tilde{T}_{25}=T_{25}$.

The ``two-point function'' with both edges on D$25$-branes
with one O$25$-bounce but without D$25$-bounces is
\begin{eqnarray}
 {1 \over {2!}}
  \langle B_{25} \vert D \hat{M}_O D\vert B_{25} \rangle
 &=& {1 \over {2!}}
     \int d^{26}x
      \langle B_{25} \vert D \vert \tilde{C}_{25}(x) \rangle
       (+\tilde{T}_{25})
      \langle \tilde{C}_{25}(x) \vert D \vert B_{25} \rangle
\nonumber\\
 &=& {1 \over {2!}} V_{26} N_{25}^2
     \hat{\Delta}_{25}
     \left( {{N_{25}} \over {T_{25}}} \right)^2 (+T_{25})
     \hat{\Delta}_{25}.
\end{eqnarray}
There are two contributions to the ``two-point function''
with both edges on D$25$-branes with one O$25$-bounce and one D$25$-bounce:
\begin{equation}
 {1 \over {2!}} {1 \over 2}
  \langle B_{25} \vert D \hat{M}_O D \hat{M} D \vert B_{25} \rangle\,
\quad
\mbox{and}
\quad
 {1 \over {2!}} {1 \over 2}
  \langle B_{25} \vert D \hat{M} D \hat{M}_O D \vert B_{25} \rangle\, ,
\end{equation}
where the overall $1/2$ is a symmetry factor, so that
\begin{eqnarray}
 &&
 {1 \over {2!}}
  \left(
   {1 \over 2}
  \langle B_{25} \vert D \hat{M}_O D \hat{M} D \vert B_{25} \rangle
    +
   {1 \over 2}
  \langle B_{25} \vert D \hat{M} D \hat{M}_O D \vert B_{25} \rangle
  \right)
\nonumber\\
 &&
  = {1 \over {2!}} V_{26} N_{25}^2
     \hat{\Delta}_{25}
     \left( {{N_{25}} \over {T_{25}}} \right)^2 (+T_{25})
     \hat{\Delta}_{25}
     \left( {{N_{25}} \over {T_{25}}} \right)^2 (-T_{25})
     \Delta_{25}.
\end{eqnarray}
These results can be understood in terms of Feynman rules for
the ``propagator'' and the ``mass insertion''.
The ``propagators'' $\langle B \vert D \vert B \rangle$ and
 $\langle C \vert D \vert C \rangle$ give $\Delta_{25}$,
 and the ``propagators''
 $\langle B \vert D \vert C \rangle$ and
 $\langle C \vert D \vert B \rangle$ give $\hat{\Delta}_{25}$.
On the other hand, the ``mass insertions'' determined by $\hat{M}$
and $\hat{M}_O$
 give $(N_{25} / T_{25})^2 (-T_{25})$
 and $(N_{25} / T_{25})^2 (+T_{25})$, respectively.
Certain symmetry factors should be included, and there is  also an
overall factor $(1/2!) V_{26} N_{25}^2$. It is then straightforward
to compute the ``two-point function'' with both edges on
D$25$-branes, with one O$25$-bounce and full D$25$-bounce:
\begin{eqnarray}
 A^{(2)}_{\rm one-O}
 &=& {1 \over {2!}} V_{26} N_{25}^2
     \hat{\Delta}_{25}
     \left( {{N_{25}} \over {T_{25}}} \right)^2 (+T_{25})
     \hat{\Delta}_{25}
\nonumber\\
 &&  \times
     \left(
     1
     +
     \left( {{N_{25}} \over {T_{25}}} \right)^2 (-T_{25})
     \Delta_{25}
     +
     \left(
      \left( {{N_{25}} \over {T_{25}}} \right)^2 (-T_{25})
      \Delta_{25}
     \right)^2
     + \cdots
     \right)
\nonumber\\
 &=& {1 \over {2!}} V_{26} N_{25}^2
     \hat{\Delta}_{25}
     \left( {{N_{25}} \over {T_{25}}} \right)^2 (+T_{25})
     \hat{\Delta}_{25}
     {1 \over {1 + T_{25} (N_{25}/T_{25})^2 \Delta_{25}}}.
\end{eqnarray}

The ``two-point functions'' with both edges on D$25$-branes with two
O$25$-bounces and full D$25$-bounce are more complicated. The
contribution without D$25$-bounces is
\begin{equation}
 {1 \over {2!}}
  \langle B_{25} \vert D \hat{M}_O D \hat{M}_O D \vert B_{25} \rangle
  = {1 \over {2!}} V_{26} N_{25}^2
    \hat{\Delta}_{25}
    \left( {{N_{25}} \over {T_{25}}} \right)^2 (+T_{25})
    \Delta_{25}
    \left( {{N_{25}} \over {T_{25}}} \right)^2 (+T_{25})
    \hat{\Delta}_{25}.
\end{equation}
The next contributions with one D$25$ bounce have three different forms:
\begin{eqnarray}
&&
{1 \over 3} {1 \over {2!}}
 \langle B_{25} \vert
  D \hat{M}_O D \hat{M}_O D \hat{M} D \vert B_{25}
 \rangle
\\
&& \qquad
 =
 {1 \over 3} {1 \over {2!}} V_{26} N_{25}^2
    \hat{\Delta}_{25}
    \left( {{N_{25}} \over {T_{25}}} \right)^2 (+T_{25})
    \Delta_{25}
    \left( {{N_{25}} \over {T_{25}}} \right)^2 (+T_{25})
    \hat{\Delta}_{25}
    \left( {{N_{25}} \over {T_{25}}} \right)^2 (-T_{25})
    \Delta_{25},
\nonumber\\
&&
 {1 \over 3} {1 \over {2!}}
 \langle B_{25} \vert
  D \hat{M}_O D \hat{M} D \hat{M}_O D \vert B_{25}
 \rangle
\\
&& \qquad
 =
 {1 \over 3} {1 \over {2!}} V_{26} N_{25}^2
    \hat{\Delta}_{25}
    \left( {{N_{25}} \over {T_{25}}} \right)^2 (+T_{25})
    \hat{\Delta}_{25}
    \left( {{N_{25}} \over {T_{25}}} \right)^2 (-T_{25})
    \hat{\Delta}_{25}
    \left( {{N_{25}} \over {T_{25}}} \right)^2 (+T_{25})
    \hat{\Delta}_{25},
\nonumber\\
&&
 {1 \over 3} {1 \over {2!}}
 \langle B_{25} \vert
  D \hat{M} D \hat{M}_O D \hat{M}_O D \vert B_{25}
 \rangle
\\
&& \qquad
 =
 {1 \over 3} {1 \over {2!}} V_{26} N_{25}^2
    \Delta_{25}
    \left( {{N_{25}} \over {T_{25}}} \right)^2 (-T_{25})
    \hat{\Delta}_{25}
    \left( {{N_{25}} \over {T_{25}}} \right)^2 (+T_{25})
    \Delta_{25}
    \left( {{N_{25}} \over {T_{25}}} \right)^2 (+T_{25})
    \hat{\Delta}_{25}.
\nonumber
\end{eqnarray}
Although the first and the third of these coincide, the second is
different. Taking only the dominant divergence due to the tachyon
tadpoles, so that $\Delta_{25} \rightarrow \Delta_T$ and
$\hat{\Delta}_{25} \rightarrow \Delta_T$, the sum of the above three
contributions becomes
\begin{equation}
 {1 \over {2!}} V_{26} N_{25}^2
    \Delta_T
    \left( {{N_{25}} \over {T_{25}}} \right)^2 (+T_{25})
    \Delta_T
    \left( {{N_{25}} \over {T_{25}}} \right)^2 (+T_{25})
    \hat{\Delta}_T
    \left( {{N_{25}} \over {T_{25}}} \right)^2 (-T_{25})
    \Delta_T.
\end{equation}
It is then straightforward to calculate the contribution from two or more D$25$-bounces,
and finally the result with full D$25$-bounce becomes
\begin{eqnarray}
 A^{(2)}_{\rm two-O}
 &=& {1 \over {2!}} V_{26} N_{25}^2
     \Delta_T
     \left( {{N_{25}} \over {T_{25}}} \right)^2 (+T_{25})
     \Delta_T
     \left( {{N_{25}} \over {T_{25}}} \right)^2 (+T_{25})
     \Delta_T
\\
 &&  \times
     \left(
     1
     +
     \left( {{N_{25}} \over {T_{25}}} \right)^2 (-T_{25})
     \Delta_T
     +
     \left(
      \left( {{N_{25}} \over {T_{25}}} \right)^2 (-T_{25})
      \Delta_T
     \right)^2
     + \cdots
     \right)
\nonumber\\
 &=& {1 \over {2!}} V_{26} N_{25}^2
     \Delta_T
     \left( {{N_{25}} \over {T_{25}}} \right)^2 (+T_{25})
     \Delta_T
     \left( {{N_{25}} \over {T_{25}}} \right)^2 (+T_{25})
     \Delta_T
     {1 \over {1 + T_{25} (N_{25}/T_{25})^2 \Delta_T}}.
\nonumber
\end{eqnarray}

It is now straightforward to obtain the contribution with full O$25$ and D$25$ bounces:
\begin{eqnarray}
 A^{(2)}_{\rm full}
 &=& {1 \over {2!}} V_{26} N_{25}^2
     {{\Delta_T} \over {1 + T_{25} (N_{25}/T_{25})^2 \Delta_T}}
\\
 &&  \times
     \left(
     1
     +
     \left( {{N_{25}} \over {T_{25}}} \right)^2 (+T_{25})
     \Delta_T
     +
     \left(
      \left( {{N_{25}} \over {T_{25}}} \right)^2 (+T_{25})
      \Delta_T
     \right)^2
     + \cdots
     \right)
\nonumber\\
 &=& {1 \over {2!}} V_{26} N_{25}^2
     {{\Delta_T} \over {1 + T_{25} (N_{25}/T_{25})^2 \Delta_T}}
     {1 \over {1 - T_{25} (N_{25}/T_{25})^2 \Delta_T}}.
\end{eqnarray}
Recalling that $\Delta_T$ is a divergent quantity, one can thus see
that the ``two-point function'' with both edges on the D$25$-brane
(cylinder with bounces), vanishes. The other two types of
``two-point functions'', with both edges on the O$25$-plane (Klein
bottle with bounces) and with one with one edge on the D$25$-brane
and the other on the O$25$-plane (M\"obius strip with bounces), are
exactly identical and thus also vanish. Since all ``two-point
function'' vanish, the ``three-point functions'' and in fact all
higher-point functions vanish, as can be simply understood by the
arguments in the previous section. In conclusion, there is no
contribution to the vacuum energy resulting from tadpole
resummation, despite the presence of the tachyon tadpole.

If we neglect the divergence introduced by tachyon tadpoles (or if
we define it by analytic continuation as $\Delta_T=-\alpha'/4$),
 letting
 $\Delta_{25} \rightarrow \Delta$ and
 $\hat{\Delta}_{25} \rightarrow - \Delta$,
 the contribution of the cylinder with bounces equals
 the contribution of the Klein bottle with bounces,
 while the contribution of cylinder with bounces equals
 $-1/2$ of the contribution of M\"obius strip with bounces.
This is easily understood,
 since the replacement of one D$25$-brane boundary state
 to one O$25$-place crosscap state gives rise to a sign change,
 due to $\Delta_{25} \rightarrow \hat{\Delta}_{25}$.
Therefore, again, all contributions to the vacuum energy arising
from tadpole resummations add up to a vanishing result, just like
the tadpole contributions, in the SO$(8192)$ theory.

\section{Conclusions}

It is interesting that these simple calculations give the result
expected by Sen's conjecture of open-string tachyon condensation
from a closed-channel perspective. The calculations combine the
boundary state formalism with some information drawn from the
low-energy effective field theory. It would be interesting to try
and extend this method to closed-string field theory, since there is
no proof that this method gives the exact result. The key problems
to be considered are the following.

In open-string tachyon condensation in String Field Theory, the
tachyon potential is obtained integrating out the open string
massive modes (see ref.~\cite{Taylor:2003gn} for a review). The
actual numerical calculation is based on the level truncation
approximation, and the obtained numerical value of the vacuum energy
is indeed very close to zero. On the other hand, the role of the
open string massive modes is not evident in our method, where
infrared divergences due to closed string tadpoles are important. It
is natural that an infinite number of open string modes are
required, since we are making explicit use of open-closed string
duality and we are tracking the tadpoles of low-lying closed string
states, the dilaton, the graviton and the closed-string tachyon.

We have accounted for the propagation of closed strings in a rigid
spacetime perpendicular to the D-branes. In particular, we have been
considering a flat spacetime, which should not be a good
approximation in absolute terms, since the very existence of
D-branes is known to lead to a back reaction on the space-time
geometry. Indeed, it was shown in ref.~\cite{Dudas:2000ff} that in
string models with broken supersymmetry without tachyons, the
dilaton tadpole curves the original flat Minkowski background,
leading to a sort of spontaneous compactification. The effect of
this gravitational back reaction is clearly not included in our
calculation,  but is similarly not included in the analyses based on
String Field Theory.
%The gravitational back reaction would introduce
%contributions of higher order in the string coupling, and we do not
%have an argument to exclude that it might alter the tadpole
%resummation.
The gravitational back reaction
 may change the contribution of the tadpole resummation.
Other subtleties resulting from the inclusion of
gravity in the tadpole resummation in field theory were discussed in
ref.~\cite{Dudas:2004nd}. It is not clear whether or not these
problems are overcome in our method.

In spite of these problems, it is straightforward to apply this
method to superstring models with ``brane supersymmetry breaking''
\cite{Sugimoto:1999tx,bsb}, \emph{i.e.} broken supersymmetry on
branes with no tachyon but dilaton tadpoles. For instance, in the
USp$(32)$ Sugimoto model \cite{Sugimoto:1999tx} all ``two-point
functions'' can be computed exactly as in SO$(8192)$ model, and the
final result is
\begin{equation}
 A^{(2)}_{\rm full}
 = {1 \over {2!}} V_{10} N_9^2
   {{\Delta_{\rm NS}} \over {1 + T_9 (N_9/T_9)^2 \Delta_{\rm NS}}}
   {1 \over {1 + T_9 (N_9/T_9)^2 \Delta_{\rm NS}}},
\end{equation}
where $N_9$ is the normalization factor of the boundary state of
D$9$-brane, $T_9$ is the tension of the D$9$-brane and $\Delta_{\rm
NS}$ includes only a divergence due to massless dilaton/graviton
tadpoles (since there is no tachyon), defined in the same way as
$\Delta_{25}$. Therefore, all ``two-point functions'' vanish, so
that there is no correction to the vacuum energy from tadpole
resummation. It would be interesting to apply this method to the
calculation of other physical quantities, and for instance to the
masses of scalar fields.

\section*{Acknowledgments}

I would like to thank Augusto Sagnotti and Emilian Dudas for
fruitful discussions and a careful reading of the manuscript. I am
especially grateful to Augusto Sagnotti for important comments and
suggestions, and for the kind hospitality during my stay at the
Scuola Normale Superiore in Pisa. This research was supported in
part by the Grant-in-Aid for Scientific Research No.~19540303 from
the Ministry of Education, Culture, Sports, Science and Technology
of Japan, and by the Italian MIUR-PRIN contract 2005-02045.

\end{document}